\documentclass[sigconf]{acmart}
\AtBeginDocument{%
  }
\usepackage{tcolorbox}
\usepackage{listings}
\usepackage{float}
\usepackage{listings}
\usepackage{xcolor}
\usepackage{graphicx}
\usepackage{placeins}
\usepackage{tcolorbox}
\usepackage{listings}
\usepackage{float}
\usepackage{xcolor}
\definecolor{lightgray}{gray}{0.9}
\usepackage{graphicx}
\usepackage{placeins}
\usepackage{booktabs} 
\usepackage{colortbl} 
\usepackage{hyperref} 
\usepackage{tabularx} 
\usepackage{enumitem}
\setcopyright{acmlicensed}
\copyrightyear{2026}
\acmYear{2026}
\acmDOI{XXXXXXX.XXXXXXX}
\acmConference[EASE 2026]{The 30th International Conference on Evaluation and Assessment in Software Engineering}{9–12 June, 2026}{Glasgow, Scotland, United Kingdom}
\acmISBN{978-1-4503-XXXX-X/2026/07}
\usepackage{booktabs} 
\usepackage{colortbl} 
\usepackage{graphicx} 
\usepackage{hyperref} 
\usepackage{tabularx} 
\usepackage{enumitem}

\definecolor{lightgray}{gray}{0.9}

\begin{document}
\setlength{\leftmargini}{1.3em}
\setlength{\leftmarginii}{1.3em}
\title{ImproBR: Bug Report Improver Using LLMs}

\author{Emre Furkan Akyol}
\affiliation{%
  \institution{Bilkent University}
  \city{Ankara}
  \country{Türkiye}
}
\email{furkan.akyol@ug.bilkent.edu.tr}

\author{Mehmet Dedeler}
\affiliation{%
  \institution{Bilkent University}
  \city{Ankara}
  \country{Türkiye}
}
\email{mehmet.dedeler@ug.bilkent.edu.tr}

\author{Eray Tüzün}
\affiliation{%
  \institution{Bilkent University}
  \city{Ankara}
  \country{Türkiye}}
\email{eraytuzun@cs.bilkent.edu.tr}

\renewcommand{\shortauthors}{Akyol et al.}

\begin{abstract}
Bug tracking systems (BTS) play a crucial role in software maintenance, yet developers frequently struggle with low-quality user-submitted reports that omit essential details such as Steps to Reproduce (S2R), Observed Behavior (OB), and Expected Behavior (EB). These inadequate descriptions lead to non-reproducible bugs, delaying resolution and wasting developer effort. 
We propose ImproBR, an LLM-based pipeline that automatically detects and improves bug reports by addressing missing, incomplete, and ambiguous S2R, OB, and EB sections. By restructuring instructions and generating clear, reproducible steps, our primary goal is to refine raw bug reports into complete, consistent, and actionable documents for developers.
ImproBR employs a multi-stage methodology centered on detection and improvement. After preprocessing raw bug reports, a hybrid detector that combines fine-tuned DistilBERT, heuristic analysis, and LLM analyzer identifies missing or ambiguous S2R, OB, and EB sections. Guided by the detector's output, ImproBR then uses GPT-4o mini with section-specific few-shot prompts to generate improvements. To enhance reliability and relevance, a Retrieval-Augmented Generation (RAG) pipeline supplements the LLM with contextual information from a knowledge base.
We evaluate ImproBR on Mojira, the bug tracker for Minecraft, a large-scale domain of user-generated, often low-quality bug reports, with reports on average only 7.9\% structurally complete, improved to an average of 96.4\% complete, by generating missing S2R, OB, and EB sections. Our manual evaluation of 139 challenging, real-world bug reports confirmed this practical impact. ImproBR more than doubled the proportion of executable S2R, from 28.8\% to 67.6\% on average, and raised the number of reproducible bug reports from just 1 to 13. 
The average run-time of our improvement pipeline for a single bug report is 23.94 seconds, helping developers avoid wasting time. These results show that ImproBR not only ensures structural completeness but also generates more semantically and procedurally accurate content for developers.

\end{abstract}


  \label{fig:teaser}


\maketitle

\section{INTRODUCTION}
\label{sec:Introduction}
Bug tracking systems (BTS) play a critical role in software maintenance by allowing developers to efficiently manage and resolve software defects. Bug reports are essential for reporting these defects from users \cite{whatmakesagood}. An essential part of a bug report is the “Steps to Reproduce” (S2R) section, which details the actions required to replicate the bug. However, in real-world software development practices, many users provide low-quality bug reports which have incomplete, ambiguous, or missing S2R sections \cite{Zhang-BugReproductionWithNLP}. As a result, low-quality bug reports force developers to spend significant time communicating back-and-forth with reporters, trying to figure out the complete reproduction path, or making assumptions, often leading to delays, increased debugging effort, and even unresolved or rejected bug reports \cite{BettenburgWhatMakesaGoodBR}. Such inefficiencies directly impact software quality, increase project costs, and lower productivity. For instance, in an empirical study conducted by Zhang et al. \cite{enrichment}, the authors analyzed Eclipse’s bug repository and found that 78.1\% of bug reports contained fewer than 100 words, and these shorter reports have an average bug resolution time of 409.8 days, that are 121 days longer than reports with 400–499 words. This evidence highlights the critical impact of low-quality bug reports on debugging efficiency and project costs. Thus, ensuring clear, structured, and actionable bug reports is crucial for efficient software maintenance processes.

Previous research on bug report quality has largely focused on detecting and classifying low-quality bug reports using various approaches, ranging from heuristic rules to machine learning classifiers \cite{chapparo-detecting-missing-informationBR,bee,euler}. Following recent advances in large language models (LLMs), researchers have begun using them to enhance bug reporting tools by automatically filling in missing or unclear sections (S2R, EB, and OB) \cite{astrobr, chatbr, adbgpt}. However, most prior work is centered on the Android domain, where reproduction and quality assessment rely on platform-specific artifacts (e.g., GUI event traces, device and OS constraints). This dependence limits generalizability and can bias evaluation toward surface-level indicators of quality. In particular, Android-oriented approaches often rely on application-specific heuristics (e.g., matching S2R to GUI components), which may improve structural completeness while still producing steps that are not executable, semantically faithful, or aligned with actual developer needs.

This study introduces \textbf{ImproBR}, an AI-powered bug report improvement pipeline aimed at enhancing low-quality user-generated bug reports by adding or improving the S2R, OB, and EB information. Our primary objective is to detect incomplete, ambiguous, and missing details by leveraging contextual and external domain knowledge to automatically produce more structured, actionable, and high-quality bug reports. To achieve this, our work focuses exclusively on more complex and testable bug report examples on the Minecraft domain, where we utilize the Minecraft Wiki\footnote{\url{https://minecraft.wiki}} as our external knowledge source. We curated our dataset from the Mojira platform \cite{mojira}, which hosts over 450,000 bug reports related to Minecraft. A significant portion of the user-generated bug reports on the Mojira platform could be considered of low quality \cite{10.1007/s10664-020-09882-z}. They often lack essential information, such as precise S2R sections or a clear description of expected behavior. 
The tangible impact of these shortcomings is further underscored by our own sampling analysis in Table \ref{tab:resolution_validation}, which demonstrated that only 11.2\% of the issues could be successfully fixed. Specifically, this work addresses the following research questions:

\textbf{RQ1:} How effectively can ImproBR improve unstructured, user-generated (raw) bug reports?

\textbf{RQ2:} To what extent does ImproBR enhance raw bug reports' semantic and contextual alignment with high-quality ground truth versions?

\textbf{RQ3:} What is the individual contribution of each ImproBR component (RAG, quality detector, and few-shot prompting) to executability and reproducibility?

Our primary contribution is a novel methodology for systematically enhancing especially low-quality bug reports. This approach leverages a detection module, combining a fine-tuned ML model, heuristics, and LLM-based analysis to identify deficiencies better in S2R, OB, and EB sections. For the improvement phase, ImproBR utilizes an LLM guided by the detector's output, section-specific few-shot prompting, and RAG. The RAG component integrates domain knowledge from the Minecraft Wiki, enhancing the semantic coherence and factual accuracy of the generated content. This multi-step approach allows ImproBR to improve the clarity, completeness, and overall quality of bug reports. 
Finally, our replication package can be found at: \url{https://figshare.com/articles/software/ImproBR_Replication_Package/30086083?file=57799795}. This package includes our manual evaluation details, semantic analysis results, the tools we used, and all source code, along with the prompts provided to the LLM.

\section{RELATED WORK} 
\label{sec:Related Work}

\subsection{Bug Report Quality Detection Tools}
\label{subsection:Bug Report Quality Detection Tools}
Since identifying the best practices for bug tracking systems are essential for efficient development, earlier studies \cite{BettenburgWhatMakesaGoodBR, bugtrackprocesssmells,burt-chatbot,whatmakesagood,enrichment, qamar2022bugtracking_smells,eray2021taxonomy} have been conducted on creating a high-quality bug report structure. These efforts focus on identifying effective bug report templates, as many users omit crucial information, such as S2R, or provide ambiguous steps, forcing developers to spend extra time querying the reporter or attempting to infer how to reproduce the issue \cite{bugtrackprocesssmells, burt-chatbot}. To address this problem, bug reports should follow a structured format that includes the system’s S2R, OB, and EB \cite{BettenburgWhatMakesaGoodBR}. 
To verify the structure of bug reports, researchers have increasingly focused on developing bug report quality detection tools that can assess and improve report structure.
For detecting and assessing the quality of bug reports, early studies focused on structural coherence and the presence of core elements. To bridge this, Bettenburg et al. \cite{BettenburgWhatMakesaGoodBR} built CUEZILLA, a prototype tool to measure bug report quality and recommend missing elements such as adding certain steps or expected results to the bug reports. To further address missing elements of the bug reports, Chaparro et al. \cite{chapparo-detecting-missing-informationBR} introduced DeMIBuD, a tool that detects whether a bug report lacks S2R, OB, and EB by leveraging heuristic NLP rules and ML-based classifiers. Similarly to this work, BEE (Bug Report Analyzer) \cite{bee} focused on extracting S2R, OB, and EB using sentence classification models, achieving accuracies of 94.7\% for OB, 99.2\% for EB, and 97.4\% for S2R, to detect missing fields and provide real-time suggestions for their completion. Lastly, EULER \cite{euler}, utilizes a neural sequence labeling model to extract structured S2R and assess their quality by matching steps with GUI components or events, including feedback on each step. Bug report improvement tools are built on top of these.
\subsection{Bug Report Improvement and Reproduction Tools}
\label{subsection:Bug Report Improvement Tools}
Several research studies have focused on reproducing bugs utilizing the S2R sections of bug reports \cite{astrobr,fusion,adbgpt,rebl,Zhang-BugReproductionWithNLP,Yapagci2025Agentic}. However, in order to achieve this, bug reports should be clear, structured, and its steps should be detailed and follow specific terminology with the application's component. As a result, automated bug report improvement is another significant challenge for researchers. Earlier studies utilized traditional NLP and ML-driven bug improvement techniques \cite{enrichment,fusion}. Recently, LLM-driven automation tools have emerged. ReBL \cite{rebl} enhances GPT’s contextual reasoning by using the entire textual bug report instead of only the S2R parts. AstroBR \cite{astrobr}, utilizes LLMs to
identify and extract the S2Rs from bug reports and map them to
GUI interactions within the program to improve S2R quality. Another LLM-based approach is AdbGPT \cite{adbgpt}, which enhances bug report understanding by extracting structured S2R entities. It maps user-described steps to device actions by leveraging entity specifications and few-shot learning. Through chain-of-thought reasoning, AdbGPT refines S2R extraction, focuses on improving accuracy in reproducing and structuring bug reports. Acharya and Ginde \cite{acharya2025enhancebugreportquality} explore instruction fine-tuning of open-source LLMs to transform unstructured bug reports into structured formats following standard templates. Their approach differs from prior work by focusing on complete bug report transformation rather than extracting specific components. They evaluate models using CTQRS, ROUGE, METEOR, and SBERT metrics, demonstrating that fine-tuned Qwen 2.5 achieves a CTQRS score of 77\%, outperforming GPT-4o in 3-shot learning (75\%). While their work demonstrates strong performance in structural transformation and field detection, the evaluation primarily emphasizes linguistic quality metrics and structural completeness.
ChatBR \cite{chatbr}, combines a fine-tuned BERT classifier with ChatGPT to assess bug reports, detect missing sections (S2R, OB, EB), and iteratively generate the absent content until a fully structured report is achieved. Their evaluation demonstrates that for detection, ChatBR improves precision by 25.38\% to 29.20\% over previous methods. For generation, ChatBR achieves an average semantic similarity of 77.62\% between the generated and original content. However, their evaluation methodology is primarily focused on the linguistic quality of the bug reports, measuring success through semantic similarity and structural completeness. Indeed, the practical value of a bug report lies in its utility for developers and QA teams. Metrics such as the reproducibility of the S2R and the factual accuracy of its terms, commands, and steps are critical evaluation criteria that are not addressed by ChatBR's semantic-focused assessment.


\section{MOTIVATING EXAMPLE}
\label{sec:Motivating Example}

To illustrate practical applicability, we applied ImproBR to three problematic bug reports from  our dataset and updated the improved versions on the official Mojira platform.

In the bug report MC-301106,\footnote{\url{https://report.bugs.mojang.com/servicedesk/customer/portal/2/MC-301106}} insufficient information was provided. This resulted in an ``Awaiting Response'' status on August 15, 2025, with developers requesting essential details such as structured reproduction steps, clear EB and OB. This report had been ``Awaiting Response'' for two months. The ImproBR-improved version of this bug report was submitted on October 14, 2025, as a comment, which includes structured S2R, clear OB, and EB.The bug that had been awaiting response for two months was reopened after our comment. On November 26, the Minecraft developers informed that this bug was fixed along with another bug report that was opened on September 5, 2025, and the issue does not exist in the latest version.

Similarly, for MC-300599,\footnote{\url{https://report.bugs.mojang.com/servicedesk/customer/portal/2/MC-300599}} was requiring additional information and had remained ``Awaiting Response'' for over three months due to unclear details. Actually, this bug was a duplicate of MC-129017\footnote{\url{https://report.bugs.mojang.com/servicedesk/customer/portal/2/MC-129017}} but could not be resolved as a duplicate of an already existing bug due to insufficient information. In contrast, when we submitted the ImproBR-enhanced version as a new report (MC-302629)\footnote{\url{https://report.bugs.mojang.com/servicedesk/customer/portal/2/MC-302629}}, it was recognized and marked as a duplicate of an existing issue within just three days.

MC-300894\footnote{\url{https://report.bugs.mojang.com/servicedesk/customer/portal/2/MC-300894}} submitted on 10 August, 2025, also contained only minimal information and the developer marked it as incomplete with "Awaiting Response" on August 11, 2025, requesting reproduction steps. The reporter, 27 days later, supplied the reproduction steps required with some in-game commands. However, the ImproBR-enhanced version had already possessed those reproduction steps, including specific in-game commands. 
 We were able to manually reproduce this bug with the ImproBR-improved version and submitted this as an additional comment.

\section{Methodology}
\label{sec:Methodology}
ImproBR's detection and improvement mechanism works as follows: A user-initiated bug report, without any additional information or comments on the original report, is provided as input to the detector, which then outputs an analysis identifying whether the report is high quality or contains missing, incomplete, or ambiguous sections that are marked as low-quality and in need of improvement. Subsequently, the low-quality bug report (preprocessed version), along with the detector’s output, is passed to the Bug Report Improver pipeline.
\begin{figure*}
    \centering
    \includegraphics[width=\linewidth]{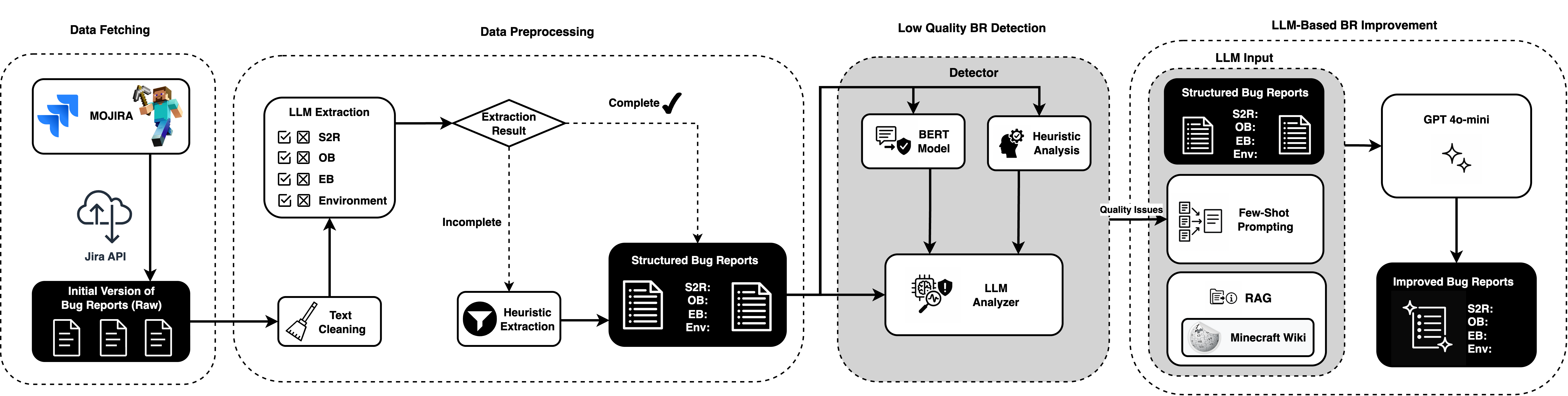}
    \caption{The ImproBR Approach}
    \Description{A flowchart showing the five stages of the ImproBR 
pipeline: data fetching, preprocessing, detection, improvement, 
and evaluation.}
\label{fig:methodology-figure}
\end{figure*}
Figure~\ref{fig:methodology-figure} illustrates the overall methodology schema and flow of our proposed approach.
Our methodology consists of five key stages (including evaluation) aimed at systematically detecting and improving missing, incomplete, and ambiguous bug reports:
Section~\ref{subsection:data-fetching} covers data fetching, 
Section~\ref{subsection:data-preprocessing} covers data preprocessing, 
Section~\ref{subsection:Low-Quality-BR-Detection} covers low-quality bug report detection, 
Section~\ref{subsection:LLM-Based Bug Report Improvement} covers bug report improvement, 
and Section~\ref{subsection:Evaluation} covers evaluation.

\subsection{Data Fetching}
\label{subsection:data-fetching}
We utilized Jira's publicly available REST API to fetch bug reports from Mojira \cite{mojira}, Minecraft's bug tracking system. Through the API endpoints, we collected essential fields including \texttt{summary}, \texttt{description}, \texttt{created}, \texttt{updated}, \texttt{status}, \texttt{resolution}, \texttt{comments}, \texttt{affected\_versions}, \texttt{priority}, and \texttt{issuelinks}.
The fetched bug reports were then forwarded to the preprocessor.
\subsection{Data Preprocessing}
\label{subsection:data-preprocessing}
To prepare our raw bug reports, which consist of a summary and an unstructured description field, we implemented a hybrid preprocessing pipeline. As depicted in Figure~\ref{fig:methodology-figure}, this pipeline extracts structured sections from unstructured text. Our approach begins with basic text cleaning to remove URLs, markdown, and HTML tags, which causes bug reports to look overcomplicated. For sentence segmentation and linguistic analysis, we employ spaCy \cite{spacy} with its \texttt{en\_core\_web\_sm} model \cite{spacy_models}, a widely adopted and efficient NLP toolkit in software engineering research.

The core of our preprocessing system is a two-level strategy. The primary extraction mechanism is GPT-4o mini \cite{openai2023gpt4o}, tasked with section identification. We use a carefully constructed few-shot prompt that instructs the LLM to be conservative, extracting content for the four key sections (S2R, Environment, OB, and EB) only when explicit headers or strong keywords are present. For example, if a user explicitly types "OB" or uses keywords commonly associated with describing observed behavior, the LLM extracts that content for the OB section. We chose GPT-4o mini for its cost efficiency and 128K token context window to process long bug reports effectively. 

For the reports where the LLM-extracted structured template is incomplete (sections are missing), our preprocessor falls back to a heuristic rule-based system to preprocess the report a second time. First, it scans for explicitly marked section headers using pattern-matching to recover content directly and integrate it into the template. If no clear headers are found, a domain-specific heuristic classifier assigns sentences to sections based on contextual clues, action verbs for S2R, version numbers for Environment, problem descriptions for OB, and normative statements for EB, supported by Minecraft-specific terminology, to direct the LLM better in ambiguous cases. Finally, we enrich the output by appending metadata (e.g., affected\_versions) into the Environment section, producing template-based reports that preserve original content while structuring it effectively for both human review and automated processing.


\subsection{Low-Quality Bug Report Detection}
\label{subsection:Low-Quality-BR-Detection}

The detection phase is designed to identify and flag ambiguous, incomplete, or missing sections within bug reports. Our approach combines three complementary methods: 

\paragraph{ML-Based Classification:}
We fine-tuned a pre-trained DistilBERT \cite{huggingface_distilbert} model on our labeled dataset to classify bug reports as low or high quality. We chose DistilBERT for its efficient balance between performance and computational requirements, as it retains 97\% of BERT's language understanding capabilities while being 40\% smaller and 60\% faster \cite{sanh2019distilbert}. We fine-tuned the entire model using cross-entropy loss, optimizing the parameters with the AdamW optimizer. We monitored its performance on a validation set to prevent overfitting. We added dataset details and the training/validation loss curves to the replication package for transparency and reproducibility.

\paragraph{Heuristic Analysis:}
To further complement our ML-based classifications, we implemented a rule-based heuristic detector that validates the structural completeness of bug reports. Our heuristic analyzer checks for three required sections: S2R, OB, and EB. For each section, it verifies their presence to correct any potential false-positive labeling errors from the BERT model.
The detector flags missing sections as quality issues. The outputs from the ML model and heuristics were combined to assign a detection output from the initial bug report, identifying critical issues related to missing sections.
\paragraph{LLM-Based Refinement:}
In parallel, we integrated GPT-4o mini into our detection pipeline as an LLM analyzer component. Our implementation uses instruction-based prompting with structured evaluation criteria. The system prompts the LLM to analyze bug reports based on specific quality dimensions, including completeness, clarity, and actionability, while explicitly focusing on their textual content. All of our prompts can be observed in the replication package.\footnote{\url{https://figshare.com/articles/software/ImproBR_Replication_Package/30086083?file=57890977}} The analyzer examines bug reports systematically, identifying not just missing sections but also the quality deficiencies such as ambiguous descriptions, inconsistent terminology, and logical gaps in reproduction steps. Ambiguity in bug reports can be referred as its existing steps cannot be reproduced due to insufficient or unrelated explanation. Sections failing these criteria are flagged by the LLM based on specificity, consistency, and coherence checks defined in our prompts.

For each report analyzed, the system outputs its judgment as a pass/fail assessment, and a detailed list of specific issues and corresponding recommendations for improvement. This evaluation incorporates previous analysis results from our ML detector and heuristics, providing the LLM with relevant context about structural and statistical assessments. By combining this structured evaluation approach with the LLM's inherent contextual understanding, our detector captures nuanced quality problems related to processed bug reports.
This integrated assessment is utilized as an input to improver LLM to determine which specific parts require enhancement in the subsequent improvement phase. Although GPT-4o-mini can detect missing or ambiguous sections, running it on every report would be too costly and slow. A separate DistilBERT detector is used for efficiency and scalability. DistilBERT provides a local filter that flags low-quality reports faster. GPT-4o-mini is then invoked only when heuristic or DistilBERT results indicate low quality, ensuring its more expensive reasoning is reserved for reports that actually require detailed refinement.

\subsection{LLM-Based Bug Report Improvement}
\label{subsection:LLM-Based Bug Report Improvement}
In the Bug Report Improvement phase, \textbf{ImproBR} combines several approaches to enhance the quality of bug reports by improving S2R, OB, and EB sections. This process involves three primary phases: Detector-Integrated improvement with GPT-4o mini, implementing RAG, and utilizing few-shot prompting techniques. By dynamically combining the detection results with domain knowledge and prompting strategies, our system provides enhancements across different bug report sections. This makes our system more reliable in the improvement phase. 
\paragraph{Detector-Integrated Improvement with GPT-4o mini:}
Our approach employs GPT-4o mini in a detector-guided pipeline. The core functionality we follow is integrating our specialized quality detection system that analyzes each bug report to identify specific deficiencies in structure, content, and completeness. These detection results are then directly fed to GPT-4o mini, providing structured guidance on which sections require improvement and what specific issues need addressing. This targeted approach enables the model to focus its capabilities on the most problematic aspects of each report while maintaining processing efficiency. 
\paragraph{Retrieval-Augmented Generation:}
To ensure high contextual accuracy, we implemented an advanced RAG pipeline \cite{lewis2020retrieval} using the LangChain framework \cite{langchain2023} and Azure OpenAI services \cite{azure2023}. Our approach begins with LLM-driven query generation, where the bug report's summary and description are analyzed to create multiple, diverse search queries. These generated queries are then used for broad candidate retrieval, fetching a wide range of documents from a Chroma vector store \cite{chromadb2023}. This knowledge base contains embeddings of official Minecraft Wiki content, generated by OpenAI's text-embedding-ada-002 model \cite{neelakantan2022text}, creating a large pool of potentially relevant information.

The most critical stage of our pipeline is Cross-Encoder Re-ranking. All retrieved candidate documents are re-ranked against the original bug report text using a sentence-transformers/ms-marco-MiniLM-L-6-v2 Cross-Encoder \cite{wang2020minilm}. The
  pipeline initially retrieves 40 candidate documents through semantic similarity search, then re-ranks and selects the top-15 most relevant
  documents, which are presented to the improvement LLM in descending relevance order to maximize contextual impact. This crucial step ensures that the knowledge ultimately selected is not just semantically similar to the search terms, but maximally relevant to the specific nuance of the original bug. As a result, our RAG pipeline lessens the risk of LLM hallucination \cite{shuster2021retrieval} and ensures the final, improved bug report is grounded in factual, domain-accurate information.
\paragraph{Few-Shot Prompting:}
To guide the GPT-4o mini model in generating structured, semantically coherent, and developer-focused bug report improvements, we implemented section-specific few-shot prompting techniques. Our approach utilizes prompts with guidelines and examples tailored to each bug report component (S2R, OB and EB). Each prompt contains contrastive examples demonstrating both poor and high-quality versions of a section. For instance, our S2R prompt includes both vague descriptions and properly enumerated steps incorporating Minecraft-specific actions.
The system first inputs each section's quality issues from the detector's output (missing, incomplete, ambiguous, or needing enhancement) and selects the appropriate prompt template based on this classification. When generating improvements, the model receives both the section-specific prompt and contextual information, including bug reports' other sections and relevant Minecraft knowledge from our retrieval system. This integrated approach helps the model to understand section-specific requirements, such as reproducibility for S2R, detailed error descriptions for Observed Behavior, and clear expected outcomes for Expected Behavior. Thus, improvements address specific deficiencies while maintaining alignment with Minecraft bug reporting conventions.

\subsection{Evaluation}
\label{subsection:Evaluation}
\begin{figure*}
    \centering
    \includegraphics[width=1.0\linewidth]{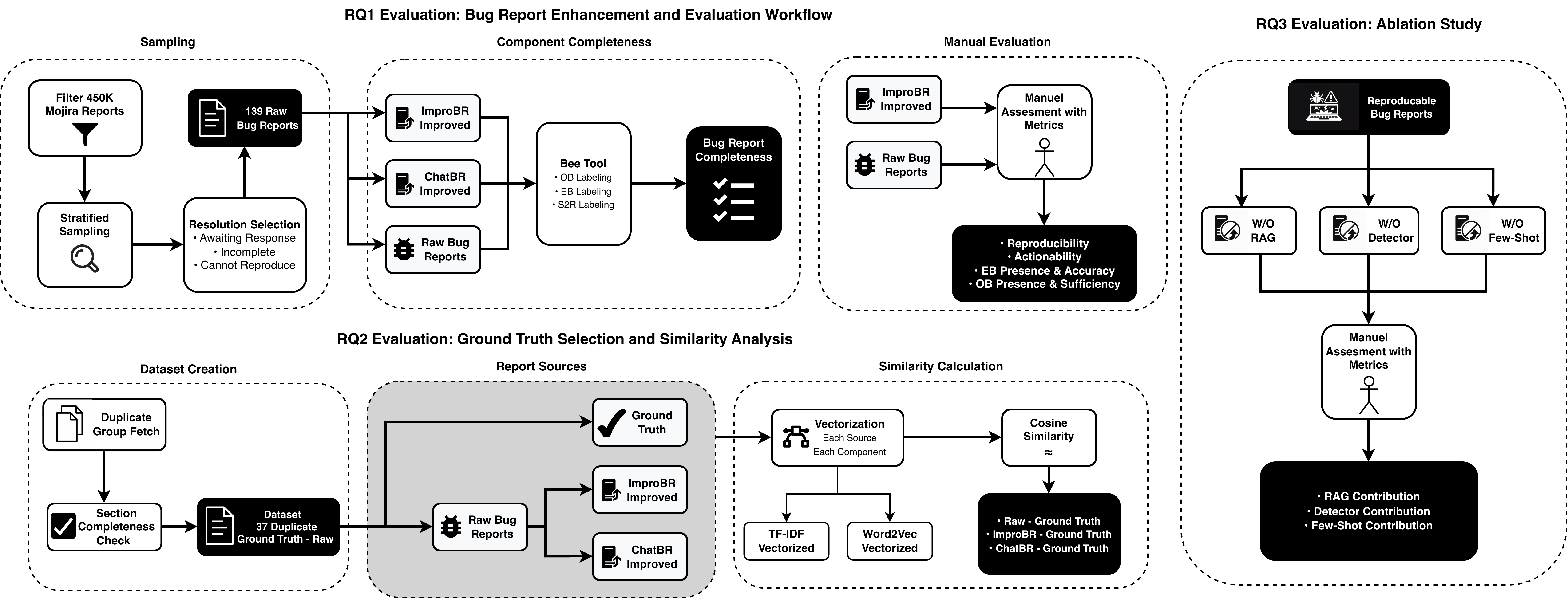}
    \caption{ImproBR Evaluation Methodology}
    \label{fig:evaluation-methodology}
\end{figure*}
To assess the effectiveness of ImproBR, our evaluation focused on quantifying the quality, reproducibility, and accuracy of the generated bug report enhancements to real-life use cases through two main research questions (RQ1 and RQ2). Figure \ref{fig:evaluation-methodology} represents the overall \textbf{ImproBR} evaluation methodology diagram.
\subsubsection{Evaluation of the Improvement by Comparing Raw and Improved Versions}

  This evaluation aimed to understand the impact of ImproBR's improvement
  process by comparing the enhanced sections against the raw report's
  preprocessed sections. 
\paragraph{Sampling Approach}
  There are nearly 450,000 bug reports available in Mojira, making it infeasible to manually evaluate every report. 
  Our sampling methodology was designed to target reports that require developer-reporter interaction and clarification. We implemented a stratified sampling approach to ensure 
  representative coverage while targeting specific resolution types. First, we retrieved a comprehensive dataset of 24,998 bug reports from Mojira using their API. We specified this limit due to the API structure change in Mojira in February 2025. By selecting our dataset from reports created after this date, we ensured a consistent bug report structure, as all reports were generated with the new API design. From this population, we 
  selected a proportional sample of 996 reports that preserved the original distribution of all 
  resolution types across the entire dataset.
Our sample size of 996 reports from a population of
  24,998 provides statistical validity with a margin of error of ±3.04\% at a 95\% confidence
  level. This means there is a 95\% probability that our observed proportions are within 3.04
  percentage points of the true population values, which is an acceptable bound for the reliability of this dataset and generalizable to the broader Mojira bug report population.

  The analysis revealed nine major resolution categories in the full dataset of 24,998 reports in Table \ref{tab:resolution_validation}. We specifically targeted three resolution types that represent cases where improved bug reports 
  could have the most impact on developer-reporter communication:   
  \begin{itemize}
      \item \textbf{Awaiting Response} (10.5\% of full dataset, 10.5\% of sample): Reports waiting for
   more information from reporter.
      \item \textbf{Cannot Reproduce} (1.8\% of full dataset, 1.8\% of sample): Reports that
  developers cannot reproduce.    \item \textbf{Incomplete} (1.6\% of full dataset, 1.6\% of sample):
  Reports with insufficient information, eventually marked incomplete.
  \end{itemize}

\begin{table}[htbp]
    \centering
    \small
    \caption{Resolution Type Distribution: Population vs. Sample}
    \label{tab:resolution_validation}
    \resizebox{0.85\linewidth}{!}{%
    \begin{tabular}{lrrrr}
    \toprule
    \textbf{Resolution Type} 
    & \multicolumn{2}{c}{\textbf{Population}} 
    & \multicolumn{2}{c}{\textbf{Sample}} \\
    \cmidrule(lr){2-3} \cmidrule(lr){4-5}
    & \textbf{(n)} & \textbf{(\%)} & \textbf{(n)} & \textbf{(\%)} \\
    \midrule
    Duplicate            & 8,720 & 34.9 & 348 & 34.9 \\
    Invalid              & 5,087 & 20.4 & 203 & 20.4 \\
    Null (Open)          & 2,862 & 11.5 & 114 & 11.4 \\
    Fixed                & 2,807 & 11.3 & 112 & 11.2 \\
    \rowcolor{lightgray}
    \textbf{Awaiting Response}  & \textbf{2,629} & \textbf{10.5} & \textbf{105} & \textbf{10.5} \\
    Works As Intended    & 1,373 &  5.5 &  54 &  5.4 \\
    Won't Fix            &   669 &  2.7 &  26 &  2.6 \\
    \rowcolor{lightgray}
    \textbf{Cannot Reproduce}   & \textbf{450}  & \textbf{1.8} & \textbf{18} & \textbf{1.8} \\
    \rowcolor{lightgray}
    \textbf{Incomplete}         & \textbf{401}  & \textbf{1.6} & \textbf{16} & \textbf{1.6} \\
    \midrule
    \textbf{Total}       & \textbf{24,998} & \textbf{100.0} & \textbf{996} & \textbf{100.0} \\
    \rowcolor{lightgray}
    \textbf{Selected for Study} & \textbf{3,480} & \textbf{13.9} & \textbf{139} & \textbf{13.9} \\
    \bottomrule
    \end{tabular}%
    }
\end{table}


  From our proportional sample, we identified all 139 reports matching these criteria (105 Awaiting 
  Response, 18 Cannot Reproduce, and 16 Incomplete), representing 13.9\% of the proportional sample. 
  These reports had an average description length of 51.6 words and 8.05 words for summary/title fields, with an average of 1.87 comments per report, potentially indicating that user-generated descriptions were characteristically problematic, which often led to developer-reporter interactions where developers stated, ‘I couldn’t reproduce this,’ etc.

  By maintaining proportional representation 
  throughout our stratified sampling process, we can draw meaningful results about ImproBR's effectiveness on the Minecraft domain
  while ensuring statistical and external validity to the broader bug reporting ecosystem.

\paragraph{Evaluation Utilizing BEE and ChatBR}
To quantitatively assess the structural completeness of our improved reports, we utilized the BEE (Bug Report Analyzer) tool \cite{bee}. BEE is a widely used framework that classifies sentences in a bug report as S2R, OB, and EB.
To ensure an objective and unbiased comparison, we used BEE as an independent, third-party detector instead of our own internal model. Our evaluation dataset consisted of the 139 raw reports curated for our manual analysis. This dataset was specifically chosen to reflect real-world scenarios where developers struggle to understand initial bug reports, thereby providing a valuable test of each model's performance on challenging use cases. We generated two sets of improved reports by processing this dataset through both the ImproBR and ChatBR pipelines. Finally, we ran the BEE tool on all three resulting datasets (raw, ImproBR, and ChatBR) to quantitatively measure the presence of the three critical elements in each version.

\subsubsection{Manual Evaluation Parameters}
Two authors independently evaluated raw and improved Mojira bug reports on the same version of Minecraft. For the S2R, authors executed each step in the reports to determine their reproducibility. After reproduction, for the OB, authors evaluated its presence and sufficiency. For the EB, authors evaluated its presence and accuracy. 

Firstly, for labeling the S2R in raw bug reports, we first followed the steps provided in the preprocessed versions, as they contained the exact same sentences from the raw bug report descriptions that had been labeled as S2R during preprocessing. If these steps alone were insufficient to reproduce the issue, we examined the raw descriptions to identify any additional, unlabeled steps that were not marked as S2R but were potentially useful for reproduction. Secondly, for labeling the S2R in improved versions, we strictly followed the steps provided. Each step cluster was then labeled according to our defined evaluation metrics, which are available in our replication package.
\footnote{ \url{https://figshare.com/articles/software/ImproBR_Replication_Package/30086083?file=57890977} \texttt{evaluation\_metrics.yaml} for detailed criteria definitions.}

\subsubsection{Improvement Evaluation by Comparing Ground truth and Improved Versions}
RQ1 aims to evaluate how ImproBR performs in improving naturally imperfect bug reports compared to ChatBR.

\paragraph{Dataset Selection}
To ensure a fair comparison, we curated 37 high-quality ground-truth bug reports and their raw duplicates from Mojira using our systematic selection algorithm. First, we modified our data fetching method to retrieve only bug reports marked as duplicates. Within each duplicate group candidate, 
we validated reports using our preprocessor and excluded those that were missing essential fields or had insufficient content. The number of remaining candidates was lower than anticipated, reflecting the prevalence of low-quality duplicates in the Mojira dataset. From the filtered set, we selected reports that contained all required sections and had substantial content as ground truths to ensure reliable similarity analysis. 
We assumed that within each duplicate group, there exists one report that developers primarily relied on to resolve the issue, and we identified this report as the ground truth. From the remaining reports in each group, excluding the selected ground-truth report, we randomly chose one as the raw version for improvement. Finally, we processed these raw reports using both the ChatBR and ImproBR pipelines for comparison.

\paragraph{ChatBR Pipeline Adaptation and Limitation}

ChatBR is a state-of-the-art method for bug report improvement, combining BERT-based classification of components (S2R, OB, EB) with LLM generation to fill in missing information \cite{chatbr}. However, its evaluation methodology has a fundamental limitation: it removes components from perfect bug reports, regenerates the missing parts, and measures success by comparing the regenerated content to the original removed text. For example, ChatBR’s pipeline removes S2R from the groundtruth EB-OB-S2R triple, feeds the remaining EB-OB to the pipeline, and measures Word2Vec similarity between the generated EB-OB-S2R triple and the original triple. The pipeline then compares the semantic similarity between the original ground truth triple and the generated triple.
However, in the real world, imperfect reports are not sliced from perfect versions. Generally, these bug reports lack well-formed components altogether, reflecting a fundamental mismatch between what developers need and what users actually provide~\cite{whatmakesagood}. Such reports often suffer from unclear descriptions, mixed component information, and incomplete details. We did not change ChatBR's prompts or fine-tuning for head-to-head comparison, preserving its full improvement mechanisms. Instead, we input the ChatBR pipeline with raw bug reports instead of removing components from ground truths, aligning with the real-world scenario. Although ChatBR’s BERT model is not included in their replication package, we retrained the same BERT on the same dataset with the same parameters, as explicitly stated in ChatBR’s replication package. These adaptations preserved ChatBR’s pipeline, ensuring a head-to-head comparison.

\paragraph{Similarity Evaluation Methodology}
After obtaining improved bug reports from both ImproBR and ChatBR, we evaluate their quality against ground truth bug reports using two complementary similarity measures. We also evaluate the raw versions before improvement to establish a baseline. Since raw reports score low against ground truth (Section~\ref{sec:Results}), a pipeline that only makes small changes to the original text cannot reach high similarity, and gains above this baseline reflect real content added by the pipeline. We use similarity scores against ground truth as the primary evaluation for this research question because ground truth bug reports represent reports that conveyed the information developers needed to understand and reproduce bugs. 
Therefore, semantic similarity against ground truth reports measures the degree to which an improved report preserves the essential information content that makes bug reports useful to developers. 
This approach is also consistent with recent bug report improvement research, where semantic similarity to original content serves as the primary quality metric~\cite{chatbr}. We used TF-IDF vectorized~\cite{Salton1975-fa} cosine similarity to measure lexical overlap, and Word2Vec vectorized~\cite{mikolov2013efficient} cosine similarity that averages all word embeddings in the bug report to capture semantic relationships. We implemented the same Word2Vec vectorized cosine similarity as ChatBR~\cite{chatbr} to ensure a fair comparison. To assess statistical significance, we employed the Wilcoxon signed-rank test due to our paired dataset and limited sample size; also, it is commonly recommended in empirical software engineering studies. ~\cite{arcuri2011practical}. We applied component-level analysis (6 tests: 3 components $\times$ 2 metrics) with Bonferroni correction ($\alpha = 0.0083$) and used Cliff's Delta ($\delta$) as its effect size measure ~\cite{romano2006appropriate}. Magnitudes interpreted as negligible, small, medium, or large. Details are available in the replication package.

\paragraph{Ablation Study}
To answer RQ3, we conducted an ablation study to evaluate each component's contribution to executability and reproducibility. We analyzed the S2R sections of the 13 reproducible bugs from RQ1, listed in Table \ref{tab:evaluation_results}. This resulted in 39 comparisons across the ablated variants. Each variant removes one component while keeping the others intact: \textbf{Without RAG} removes access to domain-specific Minecraft terminology and historical bug patterns. \textbf{Without Detector} removes guidance about which sections need improvement and what deficiencies exist. \textbf{Without Few-Shot} removes curated examples, relying solely on task instructions.

    


We restrict our analysis to reproducible bugs, since failures that cannot be reproduced by the full pipeline are unlikely to become reproducible in ablated variants that possess reduced functionality. We applied the same S2R manual evaluation metrics from RQ1 to each ablation variant. This comparison allowed us to isolate the contribution of each component and determine whether ImproBR's effectiveness stems from the synergistic combination of its components rather than any single element.
\section{RESULTS}
\label{sec:Results}
\subsection{RQ1 Results}
\subsubsection{Evaluation Utilizing BEE}

\begin{table}[htbp]
    \centering
    \caption{Bug Report Completeness Results}
    \label{tab:bee-detection}
    \resizebox{\linewidth}{!}{%
    \begin{tabular}{lccccc}
    \hline
    \textbf{Element} & \textbf{Raw} & \textbf{ImproBR} & \textbf{Imp.} & \textbf{ChatBR} & \textbf{Imp.} \\
    \hline
    Observed Behavior (OB)   & 85.6\% & 98.6\% & +13.0\% & 100.0\% & +14.4\% \\
    Expected Behavior (EB)   & 12.2\% & 96.4\% & +84.2\% & 95.0\%  & +82.8\% \\
    Steps to Reproduce (S2R) & 56.1\% & 97.1\% & +41.0\% & 98.6\%  & +42.5\% \\
    \hline
    \textbf{Complete Reports} & \textbf{7.9\%} & \textbf{96.4\%} & \textbf{+88.5\%} & \textbf{94.2\%} & \textbf{+86.3\%} \\
    \hline
    \end{tabular}%
    }
\end{table}

The results of our quantitative evaluation using the BEE tool are presented in Table \ref{tab:bee-detection}. The analysis reveals that ImproBR significantly improves the structural completeness of raw, user-submitted bug reports and demonstrates highly competitive performance against the state-of-the-art tool, ChatBR.
The most crucial finding is the increase in complete reports. While only 7.9\% of the raw reports contained all three essential elements (S2R, OB, and EB), ImproBR successfully enhanced these reports to a 96.4\% completion rate. This represents an absolute improvement of 88.5 percentage points, transforming the vast majority of incomplete reports into structurally sound documents that are ready for developer review.
The presence of EB surged from a baseline of just 12.2\% in raw reports to 96.4\% after being processed by ImproBR. This demonstrates the model's effectiveness in inferring and articulating the intended functionality, a piece of context that is very frequently omitted by users.
Furthermore, a substantial improvement was observed in the presence of S2R, which increased from 56.1\% in raw reports to 97.1\% in the improved versions. While the raw reports already had a high baseline for OB at 85.6\%, ImproBR further increased this to a near-perfect 98.6\%.

When compared to the ChatBR, ImproBR demonstrates a highly competitive and, in key areas, superior performance. ImproBR outperformed ChatBR in overall report completeness (96.4\% vs. 94.2\%) and in the EB generation (96.4\% vs. 95.0\%). While ChatBR achieved marginally higher detection rates for OB and S2R, ImproBR's ability to generate more complete, well-formed reports is a significant finding.
Notably, ChatBR's detector was fine-tuned on the same dataset provided by the BEE tool as depicted in ChatBR paper \cite{chatbr}. The fact that ImproBR still achieved a competitive or better performance without this training advantage underscores the robustness of our integrated approach to bug report improvement.

\subsubsection{Manual Evaluation}
\begin{table}[htbp]
\centering
\scriptsize
\caption{Bug Report Components' Manual Evaluation Results}
\label{tab:evaluation_results}
\begin{tabular}{|l|c|c|c|c|}
\hline
\textbf{Category} & \multicolumn{2}{c|}{\textbf{Raw Version}} & \multicolumn{2}{c|}{\textbf{Improved Version}} \\
\cline{2-5}
& \textbf{Count} & \textbf{(\%)} & \textbf{Count} & \textbf{(\%)} \\
\hline
\hline
\textbf{A. Steps to Reproduce (S2R)} & \textbf{139} & \textbf{100.0\%} & \textbf{139} & \textbf{100.0\%} \\
\hline
\quad Executable & 40 & 28.8\% & 94 & 67.6\% \\
\quad \quad $\bullet$ Reproducible & 1 & 0.7\% & 13 & 9.4\% \\
\quad \quad \quad $\circ$ Valid & 1 & 0.7\% & 5 & 3.6\% \\
\quad \quad \quad $\circ$ Invalid & 0 & 0.0\% & 8 & 5.8\% \\
\quad \quad $\bullet$ Irreproducible & 39 & 28.1\% & 81 & 58.2\% \\
\quad Non-Executable (Irreproducible) & 99 & 71.2\% & 45 & 32.4\% \\
\quad \quad $\bullet$ Ambiguous Information & 15 & 10.8\% & 8 & 5.8\% \\
\quad \quad $\bullet$ Missing Information & 80 & 57.6\% & 18 & 12.9\% \\
\quad \quad $\bullet$ Wrong Information & 4 & 2.9\% & 19 & 13.7\% \\
\hline
\textbf{B. Observed Behavior (OB)} & \textbf{139} & \textbf{100.0\%} & \textbf{139} & \textbf{100.0\%} \\
\hline
\quad Not Present & 3 & 2.2\% & 2 & 1.4\% \\
\quad Present & 136 & 97.8\% & 137 & 98.6\% \\
\quad \quad $\bullet$ Sufficient & 107 & 77.0\% & 113 & 81.3\% \\
\quad \quad $\bullet$ Insufficient & 29 & 20.9\% & 24 & 17.3\% \\
\hline
\textbf{C. Expected Behavior (EB)} & \textbf{139} & \textbf{100.0\%} & \textbf{139} & \textbf{100.0\%} \\
\hline
\quad Not Present & 98 & 70.5\% & 4 & 2.9\% \\
\quad Present & 41 & 29.5\% & 135 & 97.1\% \\
\quad \quad $\bullet$ Accurate & 37 & 26.6\% & 113 & 81.3\% \\
\quad \quad $\bullet$ Inaccurate & 4 & 2.9\% & 22 & 15.8\% \\
\hline
\end{tabular}
\end{table}
This section presents the results of our manual evaluation on 139 bug reports, directly comparing the quality of the raw bug reports against the versions enhanced by ImproBR across several metrics, as detailed in Section~\ref{subsection:Evaluation}. The finalized manual labeling outcomes, following inter-annotator agreement, for both raw and improved reports are summarized in Table~\ref{tab:evaluation_results}. To further clarify the S2R section in the table, we manually labeled the S2R sections as executable and non-executable step clusters. Among the executable clusters, some failed to trigger the bug, meaning they were irreproducible, while some successfully triggered the bug, meaning they were reproducible, which were then classified as valid or invalid.
\paragraph{Significance of the Dataset and Results} Because the evaluated set concentrates on awaiting response, cannot reproduce, and incomplete reports that are information-poor and problematic cases, any improvement here is significant since developers could not proceed without additional information. Initially, the lack of clarity was a major issue, with only 28.8\% of raw reports providing executable S2R. ImproBR improved this significantly to 67.6\%, which demonstrates its capability to generate clear and executable reproduction steps without critical missing information and ambiguities. Inadequate reporting by reporters typically records observed behavior without specifying the intended result, 
as indicated by 70.5\% of raw bug reports missing an explicit EB. ImproBR showed significant improvement in EB, ensuring 97.1\% of the improved reports contained a present EB, with 81.3\% of those being accurate, indicating ImproBR transformed ambiguous bug reports into specific, testable issues. While the OB was present in most raw reports (97.8\%), 20.9\% were insufficient. ImproBR increased the rate of sufficient OBs from 77.0\% to 81.3\%, indicating its capability to fill the missing information and make it clearer with domain relevance. Given the focus on abandoned bug reports—those typically closed as Incomplete, Cannot Reproduce, or Awaiting Response—raw reproducibility was only 0.7\%. ImproBR raised this to 9.4\%, showing that many reports otherwise destined for permanent closure became triggerable. Among these, 3.6\% were valid, demonstrating that ImproBR can revive abandoned reports into reproducible, actionable cases. Even recovering a small number of such reports is significant, as, without intervention, they would remain closed forever. The increase in invalid cases (from 0 to 8) reflects initial user misunderstandings of the bug, but this is still beneficial, since clarifying invalid reports helps developers avoid wasted effort.

\subsubsection{Inter-Rater Agreement for Each Bug Report Section}
To assess the reliability of our manual labeling process, we calculated Cohen's Kappa coefficient (\(\kappa\)) \cite{landis1977measurement} to measure inter-rater agreement between two independent annotators. The analysis was performed on 139 unique bug reports, each evaluated in both Raw and Improved versions, resulting in 278 total evaluations. The confusion matrices for S2R, OB, and EB annotations are presented in the replication package, respectively. Discrepancies between the sections individually assigned by the two authors were discussed and resolved in a follow-up meeting involving the independent Minecraft expert, who has more than 1000 hours of Minecraft playtime. Each bug report was annotated for three distinct label types: S2R Label (Executable: Reproducible/Irreproducible; Non-Executable), OB Label (Sufficient/Insufficient) and EB Label (Accurate/Inaccurate).

\paragraph{Handling of Empty Values}
In our labeling methodology, empty values carry semantic meaning. When both annotators marked a field as empty, this indicated mutual agreement that the corresponding section (e.g., EB) was not present in the bug report. These empty agreements were treated as valid concordances in Cohen's Kappa calculation. For instance, in the Raw versions, 93 out of 139 bug reports had both annotators agree that the Expected Behavior section was absent.



\paragraph{Inter-Rater Agreement Results}
Following Landis and Koch's interpretation guidelines \cite{landis1977measurement}, 
the inter-rater agreement analysis demonstrated robust reliability across all three label types. For the \textbf{S2R Label}, we achieved Cohen's Kappa values of 0.663 (85.6\% agreement) for Raw versions and 0.649 (84.9\% agreement) for Improved versions, both indicating substantial agreement between annotators. The \textbf{OB Label} showed similarly strong results with $\kappa$ = 0.642 (87.1\% agreement) for Raw versions and $\kappa$ = 0.675 (89.9\% agreement) for Improved versions. The highest agreement was observed for the \textbf{EB Label}, particularly in Raw versions where $\kappa$ = 0.801 (91.4\% agreement) indicated almost perfect agreement, while Improved versions achieved $\kappa$ = 0.698 (90.6\% agreement).

\paragraph{Overall Analysis}
In addressing RQ1, we evaluated ImproBR's performance on both structural and practical metrics. Structurally, the system demonstrated strong performance, surpassing ChatBR in complete reports. Unlike ChatBR, which relies solely on structural completeness and semantic similarity, our evaluation also assesses reproducibility through manual evaluation. This approach provides a more accurate assessment of real-world utility, as demonstrated by the practical examples in Section \ref{sec:Motivating Example}, where ImproBR-enhanced reports helped Minecraft developers resolve bugs on the Mojira platform. We deliberately selected an information-poor and problematic dataset that contains reports from categories requiring developer intervention. ImproBR more than doubled the proportion of executable S2R and increased fully reproducible bug reports from just 1 to 13, demonstrating that its enhancements have a direct, positive impact on reproducibility, not just on structural completeness. Recovering even a small number of such abandoned reports is significant, as without intervention, they would remain closed forever. The increase in invalid cases stemmed from the user's initial misunderstanding of the bug; since ImproBR converts the initial version into a reproducible one, we were able to categorize bug validity as well, which represents another important contribution to the bug report domain. However, our analysis also revealed a critical trade-off. While ImproBR successfully reduced the number of non-executable reports caused by ``Missing Information'' (from 80 to 18), it simultaneously increased the number of reports that were non-executable due to ``Wrong Information'' (from 4 to 19). This suggests that in its attempt to fill gaps, the LLM can generate incorrect details. These errors may stem from version differences in game mechanics, a misunderstanding of the bug's context, or the retrieval of outdated information from the knowledge base. In the future, we plan to implement additional guardrails to mitigate the generation of such incorrect information.

\subsection{RQ2 Results}
ImproBR's has significantly outperformed ChatBR across different metrics. Improvements over ChatBR are statistically significant across all six component-metric comparisons (Wilcoxon signed-rank, Bonferroni-corrected $\alpha = 0.0083$; all $p \leq 0.003$), with Cliff's $\delta$ ranging from 0.19 to 0.44, as summarized in Table~\ref{tab:similarity-results}.
\vspace{-2.5mm}
\begin{table}[htbp]
    \centering
    \small
    \setlength{\tabcolsep}{4pt}
    \caption{TF-IDF and Semantic Similarity Assessment Results}
    \label{tab:similarity-results}
    \begin{tabular}{@{}llccccrc@{}}
    \hline
    \textbf{Type} & \textbf{Metric} & \textbf{Raw} & \textbf{ChatBR} & \textbf{ImproBR} & \textbf{Diff.} & \textbf{\textit{p}} & \textbf{$\delta$} \\
    \hline
    TF-IDF
    & OB   & 12.0\% & 23.0\% & 30.1\% & \textbf{+7.1\%} & $<$0.001 & 0.29 \\
    & EB   & 7.3\%  & 16.6\% & 26.2\% & \textbf{+9.6\%} & $<$0.001 & 0.42 \\
    & S2R  & 9.4\%  & 26.7\% & 31.6\% & \textbf{+4.9\%} & 0.003    & 0.22 \\
    & \textit{Avg.} & \textit{9.6\%} & \textit{22.1\%} & \textit{29.3\%} & \textbf{\textit{+7.2\%}} & \textit{$<$0.001} & \textit{0.39} \\
    \hline
    W2V
    & OB   & 48.7\% & 72.7\% & 76.1\% & \textbf{+3.4\%} & $<$0.001 & 0.19 \\
    & EB   & 29.9\% & 67.5\% & 77.4\% & \textbf{+9.9\%} & $<$0.001 & 0.42 \\
    & S2R  & 25.4\% & 75.1\% & 77.7\% & \textbf{+2.6\%} & $<$0.001 & 0.30 \\
    & \textit{Avg.} & \textit{34.7\%} & \textit{71.8\%} & \textit{77.1\%} & \textbf{\textit{+5.3\%}} & \textit{$<$0.001} & \textit{0.44} \\
    \hline
    \end{tabular}
\end{table}
\vspace{-3mm}

\subsubsection{S2R Improvement}
Raw reports achieved only 9.4\% TF-IDF S2R similarity and 25.4\% semantic similarity, suggesting that users often omit reproduction steps entirely or provide unstructured reports. ImproBR substantially improved S2R quality over raw reports, raising TF-IDF similarity from 9.4\% to 31.6\% and semantic similarity from 25.4\% to 77.7\%, demonstrating its ability to transform abandoned bug reports into actionable ones found in high-quality reports. ImproBR achieved statistically significant gains over ChatBR in TF-IDF similarity by +4.9\% and semantic similarity by +2.6\%. ImproBR's S2R shows an improvement over ChatBR, generating better reproduction steps and aligning better with the ground truth.


\subsubsection{EB Improvement}
  The most significant performance gap was observed in the generation of EB, where ImproBR achieved the greatest improvement compared to ChatBR. Raw reports had only 7.3\% TF-IDF similarity and 29.9\% semantic similarity, reflecting the incompleteness and lack of terminology. ImproBR significantly improved both dimensions, reaching a TF-IDF similarity of 26.2\% (+18.9\%) and the semantic similarity of 77.4\% (+47.5\%). EB section constitutes the largest performance gap, with ImproBR outperforming ChatBR in TF-IDF similarity by +9.6\% (26.2\% vs. 16.6\%), and semantic similarity by +9.9\% (77.4\% vs. 67.5\%). These results indicate that ImproBR enriched EB with Minecraft’s intended mechanics and terminology more effectively than ChatBR.

\subsubsection{OB Improvement}

Raw bug reports showed the highest baseline quality for OB compared to other components, with 12.0\% TF-IDF similarity and 48.7\% semantic similarity, indicating users naturally describe what they observed in unstructured reports. Despite this stronger baseline, ImproBR still achieved improvements, reaching 30.1\% TF-IDF similarity (+18.1\%) and 76.1\% semantic similarity (+27.4\%). 
ImproBR outperforms ChatBR in TF-IDF similarity by +7.1\% (30.1\% vs. 23.0\%), and in semantic similarity by +3.4\% (76.1\% vs. 72.7\%). This indicates that both systems perform well when users provide observations, though ImproBR still yields closer to the ground truth.

\paragraph{Overall Analysis}

  As stated in Table \ref{tab:similarity-results}, raw bug reports showed severe deficiencies, with overall TF-IDF similarity of only 9.6\% and overall semantic similarity of 34.7\%. ImproBR substantially improved upon raw reports, achieving overall TF IDF similarity of 29.3\% (+19.7\%) and overall semantic similarity of 77.1\% (+42.4\%). ImproBR outperformed ChatBR in overall TF-IDF similarity by +7.2\% (29.3\% vs. 22.1\%), and overall semantic similarity by +5.3\% (77.1\% vs. 71.8\%). This significant gap suggests that ImproBR generates bug report components more closely aligned with ground truth, potentially due to its domain-specific knowledge integration.

\subsection{RQ3 Results}
\begin{table}[htbp]\vspace{-3mm}
    \centering
    \small
    \caption{Component Contribution to ImproBR Performance}
    \label{tab:ablation-compact}
    \begin{tabular}{ccccc}
    \hline
    \textbf{Configuration} & \textbf{Executable} & \textbf{Change} & \textbf{Reproducible} & \textbf{Change} \\
    \hline
    Raw             & 23.1\%  & -                    & 7.7\%   & - \\
    ImproBR (Full)  & 100.0\% & -     & 100.0\% & - \\
    \hline
    w/o RAG         & 53.8\%  & -46.2\%              & 46.2\%  & -53.8\% \\
    w/o Detector    & 69.2\%  & -30.8\%              & 61.5\%  & -38.5\% \\
    w/o Few-shot    & 61.5\%  & -38.5\%              & 61.5\%  & -38.5\% \\
    \hline
    \end{tabular}
\end{table}

Using the same inter-rater methodology as in RQ1, we achieved substantial agreement across all ablation variants. Across all reproducibility judgments, we observed an agreement rate of 82\% and a pooled Cohen’s $\kappa$ of 0.62, indicating substantial inter-annotator agreement. The replication package includes the confusion matrices and individual $\kappa$ scores for each ablation component. The results are shown in Table \ref{tab:ablation-compact}. Disabling RAG resulted in the largest performance drop, with executability falling to 53.8\% and reproducibility to 46.2\%. Notably, 83.3\% of failures without RAG were due to wrong information, where the LLM hallucinated incorrect domain-specific details such as invalid commands or wrong crafting recipes. Removing the Detector-Guided Improvement mechanism reduced executability to 69.2\%, with failures distributed across wrong information (50\%), ambiguous information (25\%), and missing information (25\%), indicating that detector guidance helps produce structured and accurate outputs. Similarly, disabling few-shot prompting yielded 61.5\% executability, with 60\% of failures attributed to wrong information. Two bugs (MC-300562 and MC-300599) failed across all ablated configurations, demonstrating that complex reports require the synergy of all three components. While raw bug reports exclusively failed due to missing information, ablated configurations predominantly failed due to wrong or ambiguous information, this shift confirms that ImproBR successfully addresses information gaps, but each component is necessary to ensure the generated content is accurate and unambiguous.

\section{Discussion}
\label{sec:Discussion}
\paragraph{Implications for Researchers}
Our work on ImproBR demonstrates that domain-specific bug report improvement tools can yield significant benefits. By transforming raw user submissions, we substantially increased the number of executable and reproducible bug reports, confirming the practical value of this approach. This study opens several avenues for future research. As LLMs continue to gain more sophisticated reasoning and agentic capabilities, these tools can be enhanced and generalized beyond a single domain to operate across diverse software projects. Furthermore, the high-quality, structured reports generated by systems like ImproBR can serve as a crucial input for the next stage of automation: automated bug reproduction. Future work could explore the correctness and reproducibility of a multi-agent system where an "improver" agent first refines a raw bug report, and a "reproducer" agent then uses that structured output to automatically execute the steps and verify the bug's occurrence, and assess its validity. Evaluating such a pipeline would be a critical step toward a fully automated bug resolution lifecycle.
\paragraph{Cost-Benefit Analysis}
When reports lack essential information, developers must request clarification, await responses, and cycles for re-evaluation that delay resolution. In BugCraft~\cite{Yapagci2025Agentic}, authors manually evaluated Mojira bug reports and found that 20 minutes of active effort per bug costs \$28.20 (using \$85/hour: \$64 median wage plus 30\% employer overhead~\cite{bls_ecec}), while the Mean Time To Reproduce extended to 3.41 days due to clarification overhead. ImproBR targets this gap by automatically adding missing information at submission time, operating at approximately \$0.10 per report, a significant cost advantage over manual effort (\$28.20), which potentially eliminates clarification cycles and significantly reduces time-to-resolution for incomplete bug reports.
\paragraph{Implications for Practitioners}
 By automatically enhancing reports before they are delivered to developers, ImproBR can reduce the developers' burden and decrease the time spent resolving bugs. These improvement tools can be scaled and integrated into major platforms like Jira and Github \cite{github, jira}, potentially becoming a game-changer for bug tracking systems.
Furthermore, future interactive bug reporting agents could be developed to communicate directly with users during report generation. These agents could detect missing or ambiguous statements, ask for clarification in real time, and automatically extract relevant error logs from local devices.
\section{Threats to Validity}
\label{sec:Threats to Validity}
\paragraph{Internal Validity}
A potential threat to internal validity stems from our use of the Minecraft Wiki for knowledge augmentation. The level of detail in this wiki may not be representative of documentation available for other software projects, potentially limiting the generalizability of our RAG component’s effectiveness. Another threat arises from our quantitative evaluation of structural completeness (RQ1), which relies on the BEE tool. Although BEE has been used in prior studies \cite{chatbr,bee}, its classifications serve only as a proxy for true report quality. To mitigate this limitation, we complemented the automated analysis with a manual evaluation of labeled samples. Another threat concerns the implementation fidelity is our comparison against ChatBR. To ensure a fair and direct comparison, we faithfully re-implemented their pipeline by adhering to their core design. We utilized their original dataset to train their BERT model and made no modifications to their fundamental detection and improvement logic. Our sole adaptation was a minor change to the data input stage, which was necessary to allow their pipeline to process the same set of naturally unstructured, raw bug reports as ImproBR. This approach ensures that our comparison evaluates the effectiveness of the core methodologies rather than minor implementation differences.


\paragraph{External Validity:}
The generalizability of our results is subject to several considerations. First, the study is confined to a single domain (Minecraft), which is a complex domain affected by lots of factors. However, its bug report characteristics may not reflect those of other software projects. Second, our comparative analysis against ChatBR was based on a curated set of 37 high-quality reports, selected due to the predominance of low-quality user-generated reports in Mojira. While this number provides useful insights, a larger sample would increase the robustness and generalizability of our findings. Third, our system’s reliance on a specific LLM, GPT-4o mini, means that the reported results are valid only utilizing this specific model. Finally, ImproBR currently focuses solely on textual data, excluding multimodal content such as screenshots and video links, which represents an important direction for future research.



\section{Conclusion}

\label{sec:Conclusion}
This paper introduced \textbf{ImproBR}, an AI-powered pipeline designed to address the common issue of low-quality user-generated bug reports by systematically enhancing their S2R, OB, and EB sections. Leveraging an LLM, a multi-step detection mechanism, and RAG with knowledge from the Minecraft Wiki, ImproBR aims to transform missing, incomplete, and ambiguous user-submitted reports into clear, structured, and actionable documents for developers.

Our evaluations confirm that ImproBR is highly effective at transforming low-quality bug reports into developer-ready documents. The practical impact of our system was demonstrated through a manual evaluation of 139 challenging real-world reports. The results show that ImproBR more than doubled the proportion of executable S2R, increasing it from 28.8\% to 67.6\%, and raised the number of fully reproducible bug reports from just one to 13. This improvement in actionability is also supported by a structural enhancement; automated analysis revealed that ImproBR increased report completeness from a mere 7.9\% to 96.4\%, primarily by generating missing EB and S2R sections. 
Taken together, these results demonstrate that ImproBR successfully transforms unstructured user feedback into complete, executable, and procedurally accurate information for developers.
\section{Acknowledgements}
This work was supported by TÜBİTAK (The Scientific and Technological Research Council of Turkey) under the 1001 Scientific and Technological Research Projects Funding Program, Project No. 125E371. The authors gratefully acknowledge TÜBİTAK for its support.

\bibliographystyle{ACM-Reference-Format}
\bibliography{sample-base}
\end{document}